\documentclass[preprint,showpacs,preprintnumbers,amsmath,amssymb]{revtex4-1}

\usepackage{graphicx}
\usepackage{dcolumn}
\usepackage{bm}
\usepackage[]{ams,amssymb,epsfig,color}
\newcommand{\be}{\begin{equation}}
\newcommand{\ee}{\end{equation}}
\begin{document}

\title{Spin-orbital phase synchronization in the magnetic field-driven electron dynamics
in a double quantum dot.}

\author{L. Chotorlishvili,$^{1,4}$ E.Ya. Sherman,$^{2,3}$ Z. Toklikishvili,$^{5}$, A. Komnik,$^{4}$ and J. Berakdar$^{1}$}
%

\affiliation{$^1$ Institut f\"ur Physik, Martin-Luther Universit\"at
Halle-Wittenberg, Heinrich-Damerow-Str.4, 06120 Halle, Germany\\
$^2$ Departamento de Quimica F\'isica, Universidad del Pa\'is Vasco UPV/EHU, 48080 Bilbao, Spain \\
$^3$ IKERBASQUE, Basque Foundation for Science, E-48011 Bilbao, Spain \\
$^4$  Institut f\"ur Theoretische Physik, Universit\"at Heidelberg, Philosophenweg 19, D-69120 Heidelberg,Germany\\
$^5$ Physics Department of the Tbilisi State University, Chavchavadze
av.3, 0128, Tbilisi, Georgia}

\begin{abstract}
We study the dynamics of an electron confined in a one-dimensional double quantum dot in the presence of driving external magnetic
fields. The orbital motion of the electron is coupled to the spin dynamics by spin
orbit interaction of the Dresselhaus type.
We derive an effective time-dependent Hamiltonian model for the orbital motion of the electron
and obtain a synchronization condition between the orbital and the spin dynamics. From this model
we deduce an analytical expression for the
Arnold tongue and propose an experimental scheme for realizing the synchronization of the orbital and spin dynamics.
\end{abstract}
 \maketitle

\section{Introduction}
Phase synchronization and related phenomena are among the most fascinating effects of nonlinear dynamics.
Besides the deep fundamental interest\cite{Landa,Glass,Anishchenko,Rulkov,Pikovsky,Osipov,Kozlov},
phase synchronization has a broad range of applications in chemistry \cite{Kocarev}, ecology
\cite{Peng}, astronomy \cite{Pavlov}, in the field of information transfer
using chaotic signals \cite{Fradkov}, and for the control of
high frequency electronic devices \cite{Strogatz}.
%
%
 In  nonlinear dissipative systems,
phase synchronization occurs if the frequency of the external
driving field is close to the eigenfrequency of the system. In
this case, for a certain frequency interval of
the driving field the oscillations of the nonlinear
dissipative system can be synchronized with the perturbing force.
Usually, for stronger driving fields the frequency interval for the
synchronization becomes broader and the synchronization protocol is more
efficient. The broad and growing interest in the phase synchronization calls for the analysis of new possible
realizations of this phenomenon.
A particularly interesting issue is the application of the
synchronization protocols for magnetic nanostructures,
which have rich applications \cite{Fradkov,Strogatz,Hirjibehedin,Rusponi,Mirkovic,Stroscio} and exhibit interesting nonlinear
dynamical properties that can be exploited as a testing ground for dynamical systems \cite{Chotorlishvili,Schwab}.

A principal challenge in nanoscience is to find an efficient procedure for the manipulation of the
systems states. A high level of accuracy
on the state control is required especially in such applications as
quantum computing, where a precise tailoring of the entangled
states is highly desirable \cite{Raimond,Shevchenko}. With this in mind several physical systems were
considered up to now, e.~g. Josephson junction qubits
and Rydberg atoms in a quantum cavities \cite{Raimond,Shevchenko},
ion traps \cite{Zahringer}, single molecular nanomagnets
\cite{Wernsdorfer}, and nanoelectromechanical resonators
\cite{Heinrich,Karabalin}. Among others, one of the most
promising systems are electron spins confined in two-dimensional quantum dots \cite{Valin-Rodriguez,Levitov,Rashba}
and in one-dimensional nanowires and nanowire-based quantum dots \cite{Pershin,Romano,Lu,Crisan}.
The key element of the corresponding models is
the spin orbit (SO) coupling term, which is linear in the electron momentum. Such momentum-dependent coupling offers a new way of
manipulating the spin by changing the electron momentum via a periodic electric field. This is the idea of the
electric-dipole spin resonance proposed by Rashba and Efros for the electrons confined in nanostructures on the scale of 10 nm \cite{Rashba}.
However, the external electric field can strongly affect the orbital dynamics
and thus  the system is driven out of the linear regime \cite{Khomitsky}. The nonlinearities usually result in a complex behavior of the
affected systems and their dynamics might become complicated and even unpredictable. On the other
hand, the nonlinearity may also lead to a number of interesting phenomena. In spite of the huge interest in the systems with
SO coupling, the influence of the spin dynamics on the orbital motion was not addressed yet in full detail.
With the present work we would like to bridge this gap. Our goal is to investigate the possibilities of controlling the orbital
motion of the electron via external magnetic fields acting on its spin and via the spin-orbit coupling.
That can be considered as opposite to the electric-dipole spin resonance protocol proposed by
Rashba and Efros \cite{Rashba}. We will demonstrate that: (i) by using an external driving  field one can achieve a
sufficient degree of control over the orbital motion, and (ii)  as a result, one can devise a very efficient
synchronization protocol of the orbital motion and the spin dynamics based on the application of a pulsed external
magnetic field.

\section{Theoretical model}
We consider a model system of a single electron confined in
a double quantum dot described by a potential of the form
$U(x)=U_{0}\left[-2\left({x}/{d}\right)^{2}+\left({x}/{d}\right)^{4}\right]$.
Here $U_{0}$ is the energy  barrier separating two minima with $2d$ being the distance between them.
We assume the system is dissipative, and the dissipation is a thermal effect
appearing due to a coupling to environment. The dissipation,
which  impacts mainly on the orbital motion, is essential
for the synchronization processes we are going to discuss later in the text.
For strong driving magnetic fields, the influence of the dissipation on the spin dynamics is negligibly small
and can be ignored.  In addition, we assume that the temperature is low
enough to prevent the activated over-the-barrier motion. For the particular value of
$U_{0}\sim20$ meV the low-temperature regime means $T<100$ K. For the GaAs-based structure with the electron
effective mass $m$ being 0.067 of the free electron mass and $d\sim100$ nm, the tunneling
probability is small and a classical consideration is
justified. To quantify the SO interaction we use a coupling term of the Dresselhaus type $H_{\rm so}=\alpha P_{x}\sigma^{x}$, where
$P_{x}$ is the momentum of the electron and $\sigma^{x}$ is the Pauli matrix.
Therefore, the Hamiltonian of the one dimensional system reads:
\be
H=\frac{P_{x}^{2}}{2m}+U(x)+\alpha
P_{x}\sigma^{x}+\mu_{B}gB_{z}(t)\frac{\sigma^{z}}{2}+\mu_{B}gB_{x}(t)\frac{\sigma^{x}}{2},
\label{eq:ham}
\ee
where $\mu_{B}$ is the Bohr magneton and $g$ is the electron Land\'{e} factor. Here $B_{z}(t)=B_{0}\sum\limits_{t=0}^{\infty}\delta_{\tau}(t-nT)$ is
an infinite series of external magnetic field pulses
with the pulse strength $B_{0}$, which is applied along the $z$ axis. The temporal width
of the pulses applied along the $z$-axis is smaller than the interval between the pulses $\tau\ll T$ (in what follows we set  $T=1$).
On the other hand, for the pulses along the $x$-axis
$B_{x}(t)=B_{0}\sum\limits_{t=0}^{\infty}\Delta_{T}(t-\tau n)$ the
pulse duration is larger than interval between pulses $T\ll \tau$ (cf. Fig. 1). A different route to the control
of the spin-dynamics in double quantum dots via electric field pulses is outlined in \cite{jonas}.

\begin{figure}[t]
 \centering
  \includegraphics[width=8cm]{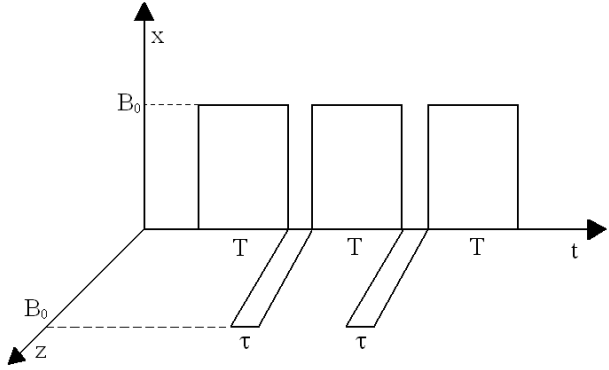}
  \caption{Schematic illustration of the infinite series of external magnetic field pulses applied to the system.
A series of the short pulses $B_{z}(t)=B_{0}\sum\limits_{t=0}^{\infty}\delta_{\tau}(t-nT)$ with the pulse width $\tau$
and time interval between pulses $T$, is applied
along the $z$ axis. Series of the pulses with the larger width $T$ and a shorter interval between the
pulses $\tau$, $B_{x}(t)=B_{0}\sum\limits_{t=0}^{\infty}\Delta_{T}(t-\tau n)$ is applied along the $x$ axis.
The amplitude of the pulses $B_{0}$ is the same in  both cases.}
\label{Fig:1}
\end{figure}

Introducing the characteristic maximum momentum of the electron $P_{x}^{\max}=\sqrt{2m U_{0}}$
we can estimate the maximal precession rate of the spin due to the SO coupling
$\Omega_{\rm so}^{\max}=({2\alpha}/{\hbar})\sqrt{2m U_{0}}$, while the magnetic field pulse of the amplitude $B=B_{0}$
induces a spin precession with the rate
$\Omega_{B}={\mu_{B}|g| B_{0}}/{\hbar}$. Therefore, if
$\Omega_{B}>\Omega_{\rm so}^{\max}$
we can  during the pulse neglect the spin rotation produced by the SO coupling and the spin is completely
controlled by the external driving fields. We need a
protocol with two driving fields in order to fulfill the
synchronization requirements as discussed later in the text. Namely, for the control of the spin
dynamics via the external driving fields, the amplitudes of the fields should be large,
$B_{0}>({2\alpha}/{\mu_{B}|g|})\sqrt{2mU_{0}}$.
On the other hand, a strong constant magnetic field
produces a high frequency precession of the spin
$\Omega_{B}={\mu_{B}|g|B_{0}}/{\hbar}$, while for
the synchronization we need to tune the precession frequency up or down keeping fixed  the strong driving field amplitude.
Below we will show  that the optimal conditions for the synchronization are realized using two types of the driving pulses.
Applying short pulses along the $z$-axis $\tau\ll T$,
$B_{z}(t)=B_{0}\sum\limits_{t=0}^{\infty}\delta_{\tau}(t-nT)$ and long pulses along the $x$-axis
$B_{x}(t)=B_{0}\sum\limits_{t=0}^{\infty}\Delta_{T}(t-\tau n)$,
we can realize a spin precession with a frequency, which is
inversely proportional to the time interval between the short pulses
$\Omega\sim 1/T$ independently from the driving field strength
$B_{0}$. In what follows, for convenience we use dimensionless units via the
transformations $E\rightarrow E/4U_{0}$, $x\rightarrow x/d$, $t\rightarrow t\sqrt{{4U_{0}}/{m}}$,
$P_x\rightarrow{P_x}/{\sqrt{2mU_{0}}}$, $\varepsilon\rightarrow\alpha/4U_{0}$.

\section{Dissipative system and the problem of phase synchronization between
orbital and spin motion}

\subsection{Spin dynamics in pulsed magnetic fields}

\begin{figure}[t]
\centering
\includegraphics[width=8cm]{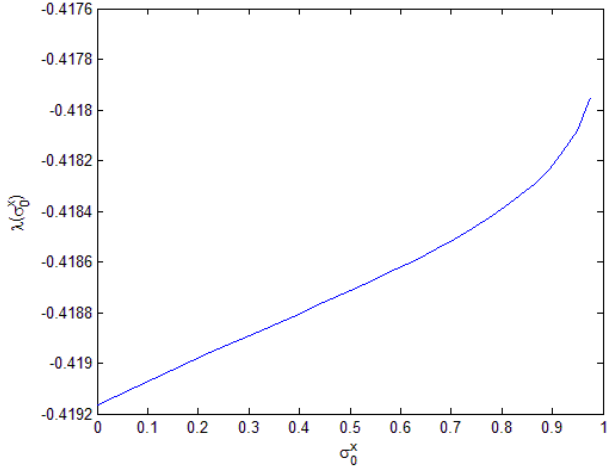}
\caption{Lyapunov exponent as a function of the initial spin component $\sigma_{0}^{x}$. After $N=1000$ iterations
the Lyapunov exponent is negative $\lambda(\sigma_{0}^{x})<0$, meaning that the spin dynamics is regular. $T=1$,
$\Omega_{B}\tau=1$, $\Omega_{B} T=20$. At these conditions, the dependence is weak.}
\label{Fig:2}
\end{figure}

\begin{figure}[t]
 \centering
  \includegraphics[width=8cm]{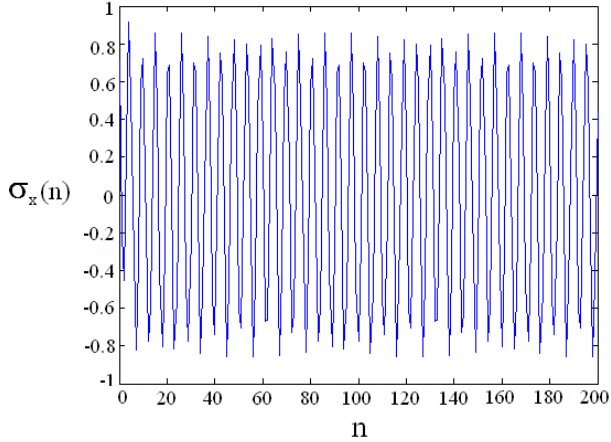}
  \caption{Time dependence of the spin projection  $\sigma_{n}^{x}\equiv\sigma^{x}(n)$. $T=1$, $\Omega_{B}\tau=1$, $\Omega_{B}T=20$.
  From Fig.~\ref{Fig:3} we see that $\sigma_{n}^{x}=\sigma^{x}(t)$ is periodic in time and the frequency of oscillations $\Omega$
  is inversely proportional to the time lapse between pulses $\Omega\approx 1/T$.} \label{Fig:3}
\end{figure}

As was stated above the synchronization can occur if the frequency of the
driving field is close to the eigenfrequency of the nonlinear dissipative system.
If this is the case, in the particular frequency interval, the oscillations of the
nonlinear dissipative system and the field can be synchronized. With the increase in the driving field
amplitude, the  synchronization  can occur in a broader frequency interval,
and the synchronization protocol becomes more efficient. Our aim is to develop a method for the synchronization of the
dynamics of the electron spin and the orbital motion, using an external driving magnetic field and SO coupling.
Although the magnetic field is not coupled to the orbital motion directly, a sufficiently strong  field influences
the orbital motion through the spin dynamics if the SO coupling is present. If SO term is relatively
small $\Omega_{\rm so}^{\max}<\Omega_{B}$, that is
\be
\Omega_{B}=\frac{\mu_{B}|g|
B_{0}}{\hbar}>\Omega_{\rm so}^{\max}=\frac{2\alpha}{\hbar}\sqrt{2mU_{0}},\label{eq:omega}
\ee
the spin and, correspondingly, the orbital motion can be controlled externally.
From Eq.~(\ref{eq:ham}) it is easy to see, that in between the short pulses
the electron spin rotates around the $x$-axis and the equations
of motion for the electron spin in this case read
\begin{equation}\label{eq:betpulse}
\dot{\sigma}^{x}=0,\quad \dot{\sigma}^{y}=-\Omega_{B}\sigma^{z}, \quad \dot{\sigma}^{z}=\Omega_{B}\sigma^{y}.
\end{equation}
On the other hand, during the short pulses we have
\begin{equation}\label{eq:shortpulse}
\dot{\sigma}^{x}=-\Omega_{B}\sigma^{y},\quad \dot{\sigma}^{y}=\Omega_{B}\sigma^{x},\quad \dot{\sigma}^{z}=0.
\end{equation}
Considering the
dynamics due to the pulse acting on the spin at the moment of
time $t=t_{0}$, we can split the evolution operator $\hat{T}_{\rm ev}$
defined  as
\begin{equation}\label{eq:Toperator:1}
\bm{\sigma}\left(t_{0}^{[-]}+T\right)=\hat{T}_{\rm ev}\bm{\sigma}\left(t_{0}^{[-]}\right)
\end{equation}
into two parts: $\hat{T}_{\rm ev}=\hat{T}_{R}\times\hat{T}_{\delta}$, where
\begin{eqnarray}\label{eq:Toperator:2}
&&\bm{\sigma}\left(t_{0}^{[+]}\right)=\hat{T}_{\delta}\bm{\sigma}\left(t_{0}^{[-]}\right),\nonumber\\
&&\bm{\sigma}\left(t_{0}^{[-]}+T\right)=\hat{T}_{R}\bm{\sigma}\left(t_{0}^{[+]}\right).
\end{eqnarray}
Here we introduced the notations $t_{0}^{[+]}\equiv t_{0}+0$ and $t_{0}^{[-]}\equiv t_{0}-0$.
The operator $\hat{T}_{R}$ describes the rotation of the
electron spin around the $x$-axis produced by the long pulse of the external driving magnetic field
$B_{x}(t)=B_{0}\sum\limits_{t=0}^{\infty}\Delta_{T}(t-\tau n)$, which is
applied along the $x$ -axis and $\hat{T}_{\sigma}$ corresponds to the
evolution produced by the short pulses
$B_{z}(t)=B_{0}\sum\limits_{t=0}^{\infty}\delta_{\tau}(t-nT)$
applied along the $z$-axis. Integrating Eq.~(\ref{eq:shortpulse})
for a short time interval $t\in\left(t_{0}^{[-]},t_{0}^{[+]}\right)$ we obtain
\begin{eqnarray}\label{eq:sigma}
&&\hat{T}_{\delta}(\sigma^{x})=\sigma^{x}(t_{0}^{[-]})\cos(\Omega_{B}\tau)-\sigma^{y}(t_{0}^{[-]})\sin(\Omega_{B}\tau),\\
&&\hat{T}_{\delta}(\sigma^{y})=\sigma^{x}(t_{0}^{[-]})\sin(\Omega_{B}\tau)+\sigma^{y}(t_{0}^{[-]})\cos(\Omega_{B}\tau).
\end{eqnarray}
 Integrating Eq.~(\ref{eq:betpulse}) during the  time interval
 $t\in\left(t_{0}^{[+]},t_{0}^{[-]}+T\right)$  of long applied pulse we find
\begin{eqnarray}\label{eq:R}
&&\hat{T}_{R}(\sigma^{y})=\sqrt{1-\left(\sigma^{x}\left(t_{0}^{[+]}\right)\right)^2}\cos(\Omega_{B}T),\\
&&\hat{T}_{R}(\sigma^{z})=\sqrt{1-\left(\sigma^{x}\left(t_{0}^{[+]}\right)\right)^2}\sin(\Omega_{B}T).
\end{eqnarray}
Combining Eq.~(\ref{eq:sigma}) with Eq.~(\ref{eq:R}) we finally  can reconstruct the complete picture of the full time evolution of the
electron spin: \be \label{eq:reccurent}\left\{\begin{array}{ll}
\sigma_{n+1}^{y}=\sqrt{1-(\sigma_{n+1}^{x})^2}\cos((n+1)\Omega_{B}T),\\
\sigma_{n+1}^{z}=\sqrt{1-(\sigma_{n+1}^{x})^2}\sin((n+1)\Omega_{B}T),\\
\sigma_{n+1}^{x}=\sigma_{n}^{x}\cos(\Omega_{B}\tau)-\sqrt{1-(\sigma_{n}^{x})^{2}}\sin(\Omega_{B}\tau)\cos(n\Omega_{B}T).\end{array}\right.\ee
The recurrent relations
Eq.~(\ref{eq:reccurent}) describe the spin dynamics. Accuracy of the employed
approximations may be checked by the validity of the normalization condition
${\bm\sigma}^{2}=1.$
In order to identify, whether the nonlinear map (\ref{eq:reccurent}) is chaotic or regular,
we evaluate the Lyapunov exponent for the spin system
\cite{Schuster}. Taking into account the peculiarity of the system (\ref{eq:reccurent}), which is the
fact, that the equation for the $x$
component $\sigma_{n+1}^{x}$ is self-consistent
$\sigma_{n+1}^{x}=f(\sigma_{n}^{x})$, we deduce for the Lyapunov exponent
\be\label{eq:liap}
\lambda(\sigma_{0}^{x})=\lim\limits_{N\rightarrow\infty \atop
\delta\sigma\rightarrow0}\frac{1}{N}\ln\left|\frac{f^{N}(\sigma_{0}^{x}+\delta\sigma)-f^{N}(\sigma_{0}^{x})}{\delta\sigma}\right|=
\lim\limits_{N\rightarrow\infty}\frac{1}{N}\ln\left|\frac{df^{N}(\sigma_{0}^{x})}{d\sigma_{0}^{x}}\right| \, .
\ee
Here, $\sigma_{0}^{x}$ is the initial value of the spin projection
and the small increment of the initial values
$\sigma_{0}^{x}+\delta\sigma$ quantifies the sensitivity of the
recurrence relations (\ref{eq:reccurent}) with respect to the
slight change in the initial conditions.
 After some algebra from Eqs.~(\ref{eq:reccurent}) and
(\ref{eq:liap}) we finally  obtain:
\be\label{eq:liapunov}
\lambda(\sigma_{0}^{x})=\lim\limits_{N\rightarrow\infty}\frac{1}{N}
\sum\limits_{n=0}^{N-1}\ln\left|\frac{df(\sigma_{n}^{x})}{d\sigma_{n}^{x}}\right|
=\frac{1}{N}\sum\limits_{n=0}^{N-1}\ln\left|\cos(\Omega_{B})+\frac{\sigma_{n}^{x}}
{\sqrt{1-(\sigma_{n}^{x})^2}}\sin(\Omega_{B})\cos(n\Omega_{B})\right| \, . \ee The results of the numerical calculations are presented on Figs.~\ref{Fig:2} and \ref{Fig:3}.

From Fig.~\ref{Fig:2} we see that the Lyapunov exponent is negative,
$\lambda(\sigma_{0}^{x})<0$  and therefore the spin
dynamics is regular, since the initial distance between two
neighboring trajectories starting from the initial points
$\sigma^{x}_{0}$ and $\sigma^{x}_{0}+\delta\sigma$ is not increasing
asymptotically after an infinite number of iterations $\delta\sigma
e^{N\lambda(\sigma_{0})}$. Therefore, from Figs.~\ref{Fig:2} and
\ref{Fig:3} we conclude, that the dynamics of the electron spin is controlled by the magnetic field pulses, thus following  equation
$\sigma^{x}(t)=\sigma^{x}_{0}\cos(\Omega t)$  we achieve the spin manipulation
by magnetic fields. The spin rotation frequency is determined by the time interval between the short
pulses $\Omega\approx 1/T$.

\subsection{Synchronization of the spin and the orbital motion.}

With the spin dynamics discussed in the previous subsection, for the orbital motion of
electron we can write the following effective Hamiltonian assuming that $\sigma^{x}(t)\approx\cos(\Omega t)$:
\begin{equation}\label{eq:model}
H=\frac{P_{x}^{2}}{2m}+U(x)+\alpha P_{x}\cos(\Omega t).
\end{equation}
The equation of motion for the system (\ref{eq:model}) has the form:
\be\label{eq:dufing} \ddot{x}+\gamma \dot{x}-x+x^3=-\beta \sin(\Omega t) \, ,
\ee
where two new dimensionless quantities are
introduced: $\beta={\alpha \Omega m}/{4U_{0}}$ and $\gamma\rightarrow\gamma/4U_{0}m$.
We seek  a solution of Eq.~(\ref{eq:dufing})
using the following ansatz
\be \label{eq:ansatz}
x(t)=\frac{1}{2}A(t)e^{i \Omega t}+\frac{1}{2}A^{*}(t)e^{-i \Omega
t} \, .
\ee
Assuming that the amplitude $A(t)$ in Eq.~(\ref{eq:ansatz}) is a slow variable the following condition applies \be\label{eq:cond}
\dot{A}(t)e^{i \Omega t}+\dot{A}^{*}(t)e^{-i \Omega t}=0
\, .
\ee
Taking into account Eqs.~(\ref{eq:ansatz}) and (\ref{eq:cond}) from Eq.~(\ref{eq:dufing}) we deduce:
\begin{eqnarray}\label{eq:amplmotion}
&& i \Omega \dot{A}(t)e^{i \Omega t}-
\left(\frac{\Omega^2}{2}A(t)e^{i\Omega t}+\mbox{c.c.}\right)+\nonumber\\
&&+\gamma\left(\frac{i \Omega}{2}A(t)e^{i\Omega t}+\mbox{c.c.}\right)-
\left(\frac{1}{2}A(t)e^{i\Omega t}++\mbox{c.c.}\right)+\\
&&+\frac{1}{8}\left(A^{3}(t)e^{3i\Omega t}+3|A(t)|^{2}A(t)e^{i\Omega t}+\mbox{c.c.}\right)=
\frac{\beta}{2i}\left(e^{i\Omega t}-\mbox{c.c.}\right)\, .\nonumber
\end{eqnarray}
\begin{figure}[t]
 \centering
  \includegraphics[width=8cm]{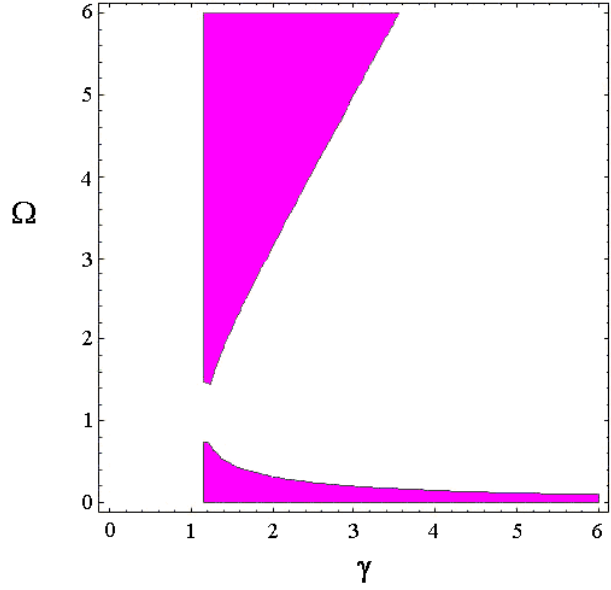}
  \caption{Arnold tongue diagram in terms of the decay constant $\gamma$ and field frequency $\Omega$,
   plotted using the conditions (\ref{eq:condition}). Shadowed regions define parameter values for which the synchronization is possible. }
  \label{Fig:4}\end{figure}
  \begin{figure}[t]
 \centering
  \includegraphics[width=8cm]{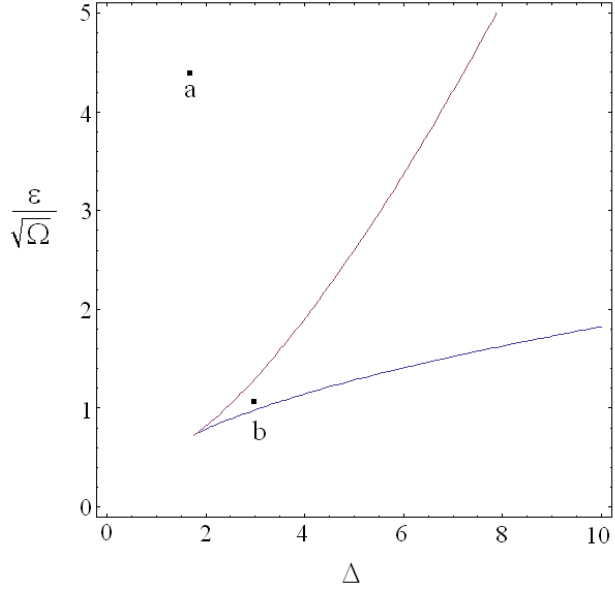}
  \caption{Arnold tongue in terms of the parameters $\Delta=\left(\Omega^2+1\right)/{\gamma\Omega}$, $(\Delta,\varepsilon)$ plotted using Eq.~(\ref{eq:parametric}).
  Point $b$  belongs to the domain where a synchronization is not possible, while $a$  belongs to the synchronization domain.}
  \label{Fig:5}
  \end{figure}
  \begin{figure}[t]
 \centering
  \includegraphics[width=16cm]{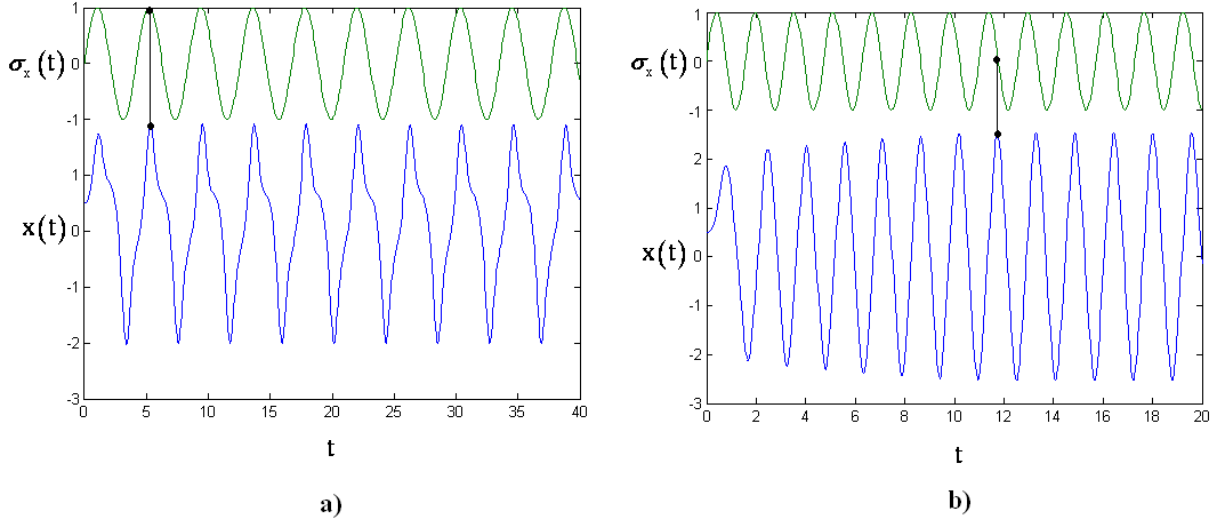}
  \caption{Orbital and spin dynamics plotted using the exact numerical integration of the Eq.~(\ref{eq:dufing}).   Panel
   a) corresponds to the values of parameters for which a synchronization is possible (see Fig.~\ref{Fig:5}, point a).
    In particular $\left({\varepsilon}/{\sqrt{\Omega}}=4.41,\Delta=1.88\right)$. The right panel b) corresponds to the point b on
    Fig.\ref{Fig:5}, $\left({\varepsilon}/{\sqrt{\Omega}}=1.05,\Delta=3.03\right)$, i.e.  the dynamics occurs
    outside of the synchronization area.
We see that in the former
     case the orbital and the spin dynamics are in phase and in the latter case the phase difference is about $\pi/2$ . }
  \label{Fig:6}\end{figure}
Multiplying Eq.~(\ref{eq:amplmotion}) by the exponent $e^{-i \Omega
t}$ and averaging it over the fast phases we find:
\be\label{eq:motionAt}\dot{A}(t)+\frac{(\Omega^2)+1)i}{2\Omega}A(t)
+\frac{\gamma}{2}A(t)-\frac{3i}{8
\Omega}|A(t)|^{2}A(t)=-\frac{\beta}{2 \Omega} .
\ee
Introducing the notations
\begin{equation}
A(t)=2\sqrt{\gamma}z(t),\quad\tau=\frac{t\gamma}{2},\quad\Delta
=\frac{\Omega^2+1}{\gamma\Omega},\quad\varepsilon=\frac{\beta}{2\Omega\gamma^{3/2}},
\end{equation}
we can rewrite Eq.~(\ref{eq:motionAt}) in a more compact form
\be\label{eq:zt} \dot{z}(\tau)+i\Delta
z(\tau)=-z(\tau)+\frac{3i}{\Omega}|z(\tau)|^2z(\tau)+\varepsilon.
\ee
Inserting $z(\tau)=R(\tau)e^{i\varphi (\tau)}$ into Eq.~(\ref{eq:zt}),
for the real and imaginary parts we obtain
\be
\label{eq:rtfit}\left\{\begin{array}{ll}
\dot{R}(\tau)=-R+\varepsilon\cos\varphi(\tau),\\
\dot{\varphi}(\tau)+\Delta=\displaystyle{\frac{3}{\Omega}}R^{2}(\tau)-\displaystyle{\frac{\varepsilon}{R(\tau)}}\sin\varphi(\tau).
\end{array}\right.
\ee
Using Eq.~(\ref{eq:rtfit}) and setting $\dot{R}=0$, $\dot{\varphi}=0$
for the stationary solutions we obtain:
\be\label{eq:fxi}
f(\xi)=\xi+\xi\left(\frac{3}{\Omega}\xi-\Delta\right)^2=\varepsilon^2,\ee
\be\label{eq:xi}
\xi_{1,2}=\frac{\Omega}{9}\left(2\Delta\pm\sqrt{\Delta^2-3}\right) \, ,
\ee
where $R^2=\xi$ and $\xi_{1,2}$ are roots of the equation
${df(\xi)}/{d\xi}=0$. In order to identify the Arnold tongue
\cite{Kuznetsov}, which shows the regions in which a
synchronization is possible, we utilize the standard condition
${df(\xi)}/{d\xi}=0$. From Eq.~(\ref{eq:xi}) it is not difficult to see that the
roots of the equation ${df(\xi)}/{d\xi}=0$ are real if
$\Delta>\sqrt{3}$. Taking into account that
$\Delta=\left(\Omega^2+1\right)/{\gamma\Omega}$ we can rewrite the inequality
in the form  $\Omega^2+1>\gamma\Omega\sqrt{3}$.  Consequently  we obtain the following criteria for the synchronization
\be \label{eq:condition}
0<\Omega<\frac{1}{2}(\gamma\sqrt{3}-\sqrt{3\gamma^2-4}),\quad\Omega>\frac{1}{2}(\gamma\sqrt{3}+\sqrt{3\gamma^2-4}),\quad\gamma>\frac{2}{\sqrt{3}}
\ee Graphical representation of the conditions (\ref{eq:condition}) is shown in Fig.~4.

Eq. (25) defines the synchronization area  in terms of the external field frequency $\Omega$ and the dissipation constant $\gamma$.
The minima points of the function $f(\xi)$, (23) does not depend on the SO coupling constant $\alpha$.
Therefore criteria (25) is independent of  the  values of the SO coupling strength as well. Nevertheless,
inserting the roots $\xi_{1,2}$  of the equation ${df(\xi)}/{d\xi}=0$ into Eq.~(\ref{eq:fxi}) one can
derive  more illustrative and precise criteria in the form of the parametrical curve:
\be\label{eq:parametric}
\frac{2}{81}(\pm3\sqrt{\Delta^2-3}+9\Delta+\Delta^3\mp\Delta^2\sqrt{\Delta^2-3})-\left(\frac{\varepsilon}{\sqrt{\Omega}}\right)^2=0 .
\ee
   The parametrical curve defined by Eq.~(\ref{eq:parametric}) represents the border of the synchronization domain, see Fig.~\ref{Fig:5}.
  Taking into account that $\beta={\alpha\Omega m}/{4U_{0}}$,
  $\varepsilon={\beta}/{2\Omega\gamma^{3/2}}$ we obtain
  \be\label{eq:epsilondelta}
  \frac{\varepsilon}{\sqrt{\Omega}}=\frac{\alpha}{\sqrt{\Omega}}\frac{m}{8U_{0}\gamma^{3/2}}.
  \ee
  From Eq.~(\ref{eq:epsilondelta}) we see that the parameters of Fig.~\ref{Fig:5} depend on the oscillation frequency that can be easily controlled by
tuning the time interval between pulses $\Omega\approx 1/T$.
  All other parameters in Eq.~(\ref{eq:epsilondelta}), such as the SO coupling constant $\alpha$, barrier height $U_{0}$, and the
  electron effective mass $m$ are internal characteristics of the system whereas the decay constant  $\gamma$ is related to the thermal effects.
  Using Eq.~(\ref{eq:parametric}) and Fig.~\ref{Fig:5}
one can synchronize the electron orbital motion with its spin dynamics.

\section{Conclusions.}
We have investigated the classical electron dynamics  in a double dot potential, with the spin of electron being controlled by
external magnetic fields. We have shown that the orbital electron dynamics can  be controlled very effectively
by the field in the presence of a spin-orbit coupling. Using the proposed protocol of magnetic field pulses of different duration
we have shown that it is possible to
synchronize the spin and the orbital motion of the electron. In particular, if the driving field amplitude is large enough,
$B_{0}>({2\alpha}/{\mu_{B}|g|})\sqrt{2m U_{0}}$, the spin dynamics is periodical in time. Then $\sigma^{x}(t)=\sigma^{x}_{0}\cos(\Omega t)$, where the
oscillation frequency is inversely proportional to the time
interval between pulses $\Omega \approx 1/T$  and can be tuned
independently from the amplitude of the pulses $B_{0}$. As a consequence the orbital dynamics can be studied with reduced
effective, time-dependent, one-dimensional model. By using this model we found the
synchronization condition between the orbital and the spin dynamics. Furthermore, we derived an analytical expression for the Arnold tongue
that defines the values of the parameters for which a synchronization is possible.
Since in the designed protocol the spin precession rate is determined by the interval between
the applied pulses we believe that it can be realized in
future  experiments on semiconductor quantum dot devices.

\textbf{Acknowledgments} The
financial support by the Deutsche Forschungsgemeinschaft (DFG)
through SFB 762, and contract BE 2161/5-1, Grant No. KO-2235/3, and
STCU Grant No. 5053 is gratefully acknowledged. EYS acknowledges support of the
MCI of Spain grant FIS2009-12773-C02-01, and "Grupos Consolidados UPV/EHU
del Gobierno Vasco" grant IT-472-10.

\end{document}